\documentstyle[aps,psfig]{revtex}

\newcommand{\be}{\begin{equation}}
\newcommand{\ee}{\end{equation}}
\newcommand{\bex}{\begin{eqnarray}}
\newcommand{\eex}{\end{eqnarray}}
\begin{document}

\title{An extended Einstein-Podolsky-Rosen thought-experiment}
\author{R. Srikanth\thanks{e-mail: srik@iiap.ernet.in}}
\address{Indian Institute of Astrophysics, Koramangala, 
Bangalore- 34, Karnataka, India.}
\maketitle
\date{}


\begin{abstract}
We study a
generalization of the original Einstein-Podolsky-Rosen thought experiment.
It is essentially a delayed choice experiment applied
to entangled particles. The basic idea is: given two observers sharing 
position-momentum entangled photons, one party chooses whether she measures 
position or momentum of her photons after the particles leave the source. The 
other party should infer her action by checking for the absence or 
presence of characteristic interference patterns after subjecting his particles 
to certain optical pre-processing. 
An occurance of apparent signaling is attributed to the difficulty in treating
single photons simultaneously at a quantum mechanical and quantum 
electrodynamic level, as required by the experiment, and points to the need for
a careful study of an aspect of the foundations of quantum mechanics
and electrodynamics.
\end{abstract}
 
\section{Introduction}

Quantum information has opened up a new and exciting era in recent times both 
in fundamental and applied physics. Its nonclassical resources of quantum
superposition and entanglement are at the heart of powerful future applications
in communication \cite{ben93,dens}, computation \cite{pre98} and
cryptographic key distribution \cite{ekert}, among others.
And yet, the fundamentally very important question whether the correlated
measurements on entangled systems imply a nonlocal transfer of information
remains somehow unclear. Einstein, Podolsky and Rosen (EPR) thought that they
did, which was the basis of their claim of quantum mechanical incompleteness
\cite{epr}. Quantum nonlocal correlations have been confirmed in
experiments since the mid-1980's performed on both systems
entangled in spin and continuous variables 
(Refs. \cite{exp} and references therein). 

Bell's celebrated theorem \cite{bella} tells us
only that any realistic model of quantum mechanics should be nonlocal.
Informed opinions diverge between on the one hand the view that quantum
nonlocality implies
no information transfer, but only a change in the mutual knowledge of the two
nonlocal systems, to the acknowledgement on the other hand of a tension
between quantum theory and special relativity \cite{weihs}. The tension stems
from the possibility that the nonlocal correlations might imply a 
superluminal transfer of information. A majority of physicists in the field, 
it would seem, accept the scenario of a spacelike but causal enforcement of
correlation, as for example in quantum dense coding \cite{dens}.
In this view, entanglement cannot be used to transmit classical signals
nonlocally  
because statistically the single particle outcome at any one particle is not 
affected by measurements on its entangled twins
\cite{nosig}. This understanding is echoed in statements of 
``a deep mystery" \cite{ghz}, and ``peaceful coexistence" \cite{shi89}
between quantum nonlocality and special relativity. 

In the present article, we propose a generalization of the original
EPR Gedanken experiment to facilitate a clearer analysis of the information
transfer question, and discuss its implications for the foundations of
quantum mechanics/electrodynamics.

\section{Thought experiment}\label{thought}

The proposed thought experiment, a modification of the original
Einstein-Podolsky-Rosen experiment \cite{epr}, involves source S of light 
consisting of entangled photon-pairs, analyzed by two observers, Alice and Bob,
 spatially seperated by distance $D$, and equidistant from S.
Alice is equipped with a device to measure
momentum or position of the photons. Bob is
equipped with a Young's double-slit interferometer 
with the slits seperated along the $y$-axis. The optics in front of the
interferometer ensures that only a narrow bundle of monochromatic rays 
making angle $\theta$ to the $x$-axis
is interferometrically analyzed, the rest being deflected away
 (Figure \ref{bir}). It
consists of two convex lenses of identical focal length $f$ and aperture
radius $R$, in tandem such that their foci coincide. A reflecting shield, 
surrounding lens 1, allows only light incident on lens 1 to pass into Bob's 
apparatus. A diaphragm with a small hole of radius $h$ and centered on the 
lens axis is placed at the mutual focal plane of the lenses. This system is
oriented such that the principal axis of the lenses makes angle $\theta$
with $x$-axis.

By geometrical optics, rays parallel to $x$-axis
incident on lens 1 fall at a distance $\eta = f\theta$ from the hole 
on the diaphragm and are blocked. 
Plane waves falling on lens 1 at angles larger than
$h/f$ do not converge at the hole are absorbed by the diaphragm.
Provided $h/f \ll \lambda_0 /s$, where $s$ is the seperation between
the two slits on Bob's interferometer and $\lambda_0$ the wavelength of the
light incident on the double-slit, the angular tolerance
will not greatly affect the visibility of the fringes on Bob's screen.
Furthermore, the diffraction effect at the lens edges can be ignored provided
$R \gg \lambda_0$. Therefore, given $hs/f \ll \lambda_0 \ll R$, 
the direction filter thus built up is effective to good approximation. 
Except for the practical requirement of keeping the integration time low,
the hole size $h$ can be made arbitrarily small.
A spectral filter behind lens 2 restricts the light to a narrow bandwidth about
wavelength $\lambda_0$. 
			
At time $t_1$,  part of the light (denoted A) from S is analyzed by Alice  
while its EPR counterpart (denoted B) is incident upon Bob's lens 1.
Bob completes his observation of the interference pattern on his
screen at $t_2 \equiv t_1 + \delta t$. By prior agreement, 
Alice measures either the transverse position and longitudinal momentum 
($\hat{y}, \hat{z}, \hat{p}_x$) of the photons in ensemble A, or the momentum 
componants ($\hat{p}_x, \hat{p}_y, \hat{p}_z$), 
while Bob will observe the resulting 
interference pattern, if any, on his screen.

\subsection{Alice measures momentum}

The source is prepared so that just before Alice's measurement at
time $t_1$ the pure state of the entangled EPR 
pairs is described by the wavefunction:
\begin{equation}
\Psi ({\bf x}_a, {\bf x}_b) = \int\int\int
e^{(2\pi i/\hbar )({\bf x}_a - {\bf x}_b + {\bf x_0})\cdot{\bf p}}dp_xdp_ydp_z,
\label{wavefunc1}
\end{equation}
where ${\bf x} \equiv (x, y, z)$ is position, the
subscripts $a$ and $b$ refer to the particles in ensembles A and B,
respectively, and ${\bf p} \equiv (p_x, p_y, p_z)$ is momentum.
Here integrals with unspecified limits are taken between $-\infty$ and 
$\infty$.  Eq. (\ref{wavefunc1}) can be rewritten:
\be
\Psi ({\bf x}_a, {\bf x}_b) = \int\int\int
\pi_a({\bf x}_a)\pi_b({\bf x}_b)dp_xdp_ydp_z.
\ee
where $\pi_a({\bf x}_a) \equiv e^{(2\pi i/\hbar){\bf x_a}\cdot{\bf p}}$ and
$\pi_b({\bf x}_b) \equiv e^{(2\pi i/\hbar)({\bf x}_0 - {\bf x}_b)\cdot{\bf p}}$
are eigenfunctions corresponding, respectively, to a monochromatic
plane wavefront moving in the $+{\bf p}$ and $-{\bf p}$ directions. 

If Alice observes $(\hat{p}_x, \hat{p}_y, \hat{p}_z)$, she produces a mixed 
state wherein each photon in A has
collapsed to a momentum eigenstate $\pi_a({\bf x}_a)$, which is a plane wave
with the eigenvalue ${\bf p} = (p_x, p_y, p_z)$. 
Simultaneously, its twin photon in B is left in the eigenstate
$\pi_b({\bf x}_b)$ with momentum eigenvalue $-{\bf p}$. The actual 
value of $(p_x, p_y, p_z)$ is in general different for different photons in A, 
giving rise to a mixed state in B, too.

Bob's optics ensures that only
plane wavefronts normal to the lenses' axis incident upon lens 1 
and whose frequency falls within the allowed narrowband about
$\lambda_0$ have non-vanishing amplitude to fall on the double slit diaphragm. 
In other words, any particle detection at Bob's screen could have come
only from a narrowband of wavefronts propagating with an orientation $\theta$
and incident on lens 1.  Therefore, all ``hits" on
Bob's screen are coincidences corresponding to narrow spectral and
angular ranges in A measurements. Hence, provided the ensembles A and B 
are sufficiently large, a Young's double slit 
interference pattern forms on Bob's screen in the single counts.

\subsection{Alice measures position}

Eq. (\ref{wavefunc1}) can be expanded in the position-momentum (mixed) 
bases as:
\be
\Psi ({\bf x}_a, {\bf x}_b) = 
	\int\int\int \xi_a({\bf x}_a)\xi_b({\bf x}_b)dydzdp_x,
\ee
where $\xi_a({\bf x}_a) = \delta (y_a - y)\delta (z_a - z) 
e^{(2\pi i/\hbar )(x_a p_x)}$, which represents a monochromatic photon wave 
ray in A moving with momentum ${\bf p}^{\prime} = (p_x, 0, 0)$ being 
transversely localized in the $yz$ plane at $(y_a, z_a) = (y, z)$. And we find:
\be
\label{xidef}
\xi_b({\bf x}_b) = \hbar\delta (y - y_b + y_0)\delta (z - z_b + z_0)
e^{(2\pi i/\hbar )(x_0 - x_b)p_x},
\ee
which represents a monochromatic photon wave ray in B moving with momentum
$-{\bf p}^{\prime}$ being transversely localized at 
$(y_b, z_b) = (y + y_0, z + z_0)$. 

If Alice measures $(\hat{y}, \hat{z}, \hat{p}_x)$ on A, she produces a mixed 
state
wherein each photon has collapsed to some eigenfunction $\xi_a({\bf x}_a)$,
which can be visualized as a horizontal momentum ray with transverse
localization. Simultaneously, each twin photon in B, which is left in 
the state $\xi_b({\bf x}_b)$ and eigenvalues $(y + y_0, z + z_0)$.
Where $(y + y_0, z + z_0)$ falls outside the aperture of lens 1, the rays
are reflected off Bob's apparatus by the enclosure. Where the rays fall on
his aperture, being horizontal, they converge to a point at distance
$\eta$ from the hole, to be blocked by
the diaphragm. Hence Bob will not observe any photon on his screen. 

The preceding experiment suggests that Alice can transmit a classical 
binary signal by choosing to measure $(\hat{y}, \hat{z}, \hat{p}_x)$
or $(\hat{p}_x, \hat{p}_y, \hat{p}_z)$ on A. 
In the following section, we present specific and general arguments that
might be expected to render such a signaling system unfeasible. The above
experiment is similar to the delayed choice experiment \cite{wheeler}, except
that it involves an entangled pair rather than a single photon, with the 
effect of Alice's choice 
on Bob's, rather than her own, photon being considered.  

\section{Discussion} 

It is generally agreed that classical signal transmission via nonlocality
is not possible because of the `no-signaling' condition, according to which
Bob's probability of measurement of any eigenstate is unaffected by Alice's
action \cite{nosig}. 
This prompts the question: what is the origin of the 
nonlocal classical signal in the above thought-experiment? 

By way of preliminary response, first we note a disparity forced on us.
In the thought-experiment, the photon is analyzed in two distinct ways before
and after its incidence on the lens. The wave field {\em before} the lens, 
described by Eqs. (\ref{wavefunc1}) through (\ref{xidef}), is the unphysical
(i.e., unobservable) quantum mechanical probability wave.
On the other hand, the wave field {\em after} the lens is the measurable 
quantized radiation field and governed by quantum electrodynamics (QED). The
consistency of the experimental prediction is based on the assumption
that there is a continuous passage from the incident wave field to that
transmitted beyond the lens.
However, in fact no rigorous proof exists that this reasonable
assumption is justified-- a difficulty modern quantum formalism should 
contend with \cite{despagnat}. 

One point to be taken into consideration in evaluating the assumption is that
QED is genuinely covariant, whereas in quantum mechanics, as is well known,
spatial coordinates are operators, but time is merely a parameter. 
As position and momentum don't commute,
phase space is not well represented in quantum mechanics. On the other hand, in
QED, spatial coordinates are also parameters like time, and not operators, 
which are furnished by the fields. It remains a difficulty how the quantum
mechanical operator algebra may be recovered as an ``effective" theory in
the one-particle non-relativisitic limit of QED. 

For example, it is not position, but phase, that the quantum electrodynamics
momentum operator $\hat{\bf P} \equiv 
\sum_{k,\alpha} \hbar{\bf k}\hat{N}_{k,\alpha}$ does not commute with.
Here $\hat{N}_{k, \alpha}$ is number operator for mode $k$ with polarization
$\alpha$.  It is possible to derive a relation akin to 
the familiar Heisenberg uncertainty principle $\Delta p_x\Delta x \ge
\hbar$, based on the number-phase non-commutation relation 
(for all $k, \alpha$) 
$[\hat{N}, \hat{\phi}] = i$, but this is meaningful only as applied to light 
{\em beams}, and not
to an individual photon, whose localization it actually seems to prohibit
(since the photon number becomes highly indeterminate if phase relations
between modes are definite enough to localize the radiation field).
 
\begin{figure}
\centerline{\psfig{file=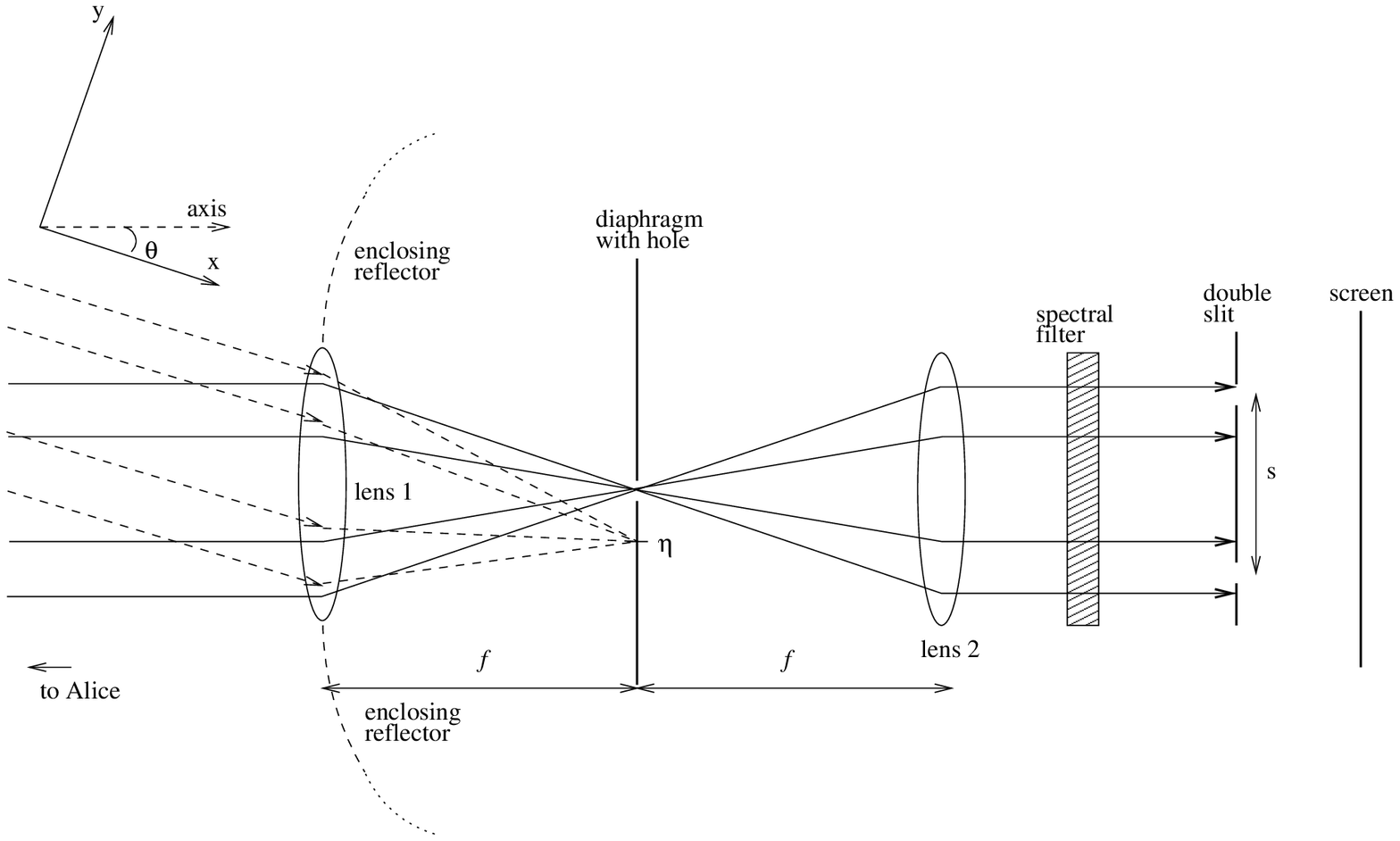,width=14.0cm}}
\caption{Bob's equipment: Its double-lens-and-reflector system acts as a 
direction filter that permits only rays parallel to the principal
axis of the lens, which is inclined at angle $\theta$ to the $x$-axis, 
to fall on the
double slit. The spectral filter restricts the light from lens 2 to a narrow 
bandwidth about $\lambda_0$. The enclosure with a reflecting exterior ensures 
that the only light entering the experiment is via lens 1. The lenses' size is
large compared to $\lambda_0$ in order to minimize correlation loss  
produced by diffraction at the lens edges. A 
$(\hat{p}_x, \hat{p}_y, \hat{p}_z)$-measurement by
Alice has a non-vanishing chance of producing an interference pattern on Bob's
screen. But $(\hat{y}, \hat{z}, \hat{p}_x)$ leaves Bob's particles as rays 
parallel to the 
$x$-axis. They converge to points $\eta$ and do not enter the interferometer.}
\label{bir}
\end{figure}

Hence, the lens in Figure \ref{bir} ends up acting as an interface between 
two disparate wave fields, whose consistency is not established in the
relevant limit.
From this viewpoint, we conclude that the claimed signaling finds its origin
in the lack of a seamless union between the applications of the two quantum 
formalisms 
and that resolving this disparity is expected to reinstate no-signaling.
 
Following a different line of thought, one might consider possible
quantum information theoretic principles that might be brought to bear upon
the experiment. It is an intriguing observation in quantum information that 
the ingredients that enforce no-signaling,
linearity \cite{gisin},  unitarity (as evidenced by no-cloning \cite{woo82})
and the tensor product character of the Hilbert space of composite quantum 
systems \cite{pikok} 
are essentially non-relativistic. It might be conjectured that
there lies a deeper connection
between quantum information and the causal structure of spacetime whereby
(e.g.,) the entangled light in the experiment somehow decoheres into 
seperable momentum pointer states \cite{zeh70}
before reaching Bob's double-slit, so that he will always find a Young's
double-slit interference pattern irrespective of Alice's action. 

Whether this, or a re-examination of the quantum formalism, suggested earlier,
or any other uncovered general principles are permitted by Nature to resolve 
the problem posed by the gedanken experiment, only suitable practical 
experiments (e.g., \cite{srik01,alt}) can adjudicate. 
The gedanken experiment shows us how  
quantum optics/information can shed light on basic issues 
in quantum theory.

\acknowledgements 
I am thankful to Dr. R. Tumulka, Dr. R. Plaga, Dr. M. Steiner, 
Dr. J. Finkelstein,  Dr. C. S. Unnikrishnan and Ms. Regina Jorgenson
for constructive criticisms and suggestions.

\newpage

\end{document}